\documentclass[prl,onecolumn,showpacs,preprintnumbers,amsmath,amssymb,preprint]{revtex4}

\usepackage{graphicx}

\usepackage{graphicx}
\usepackage{dcolumn}
\usepackage{bm}
\usepackage[latin1]{inputenc}          
\usepackage[english]{babel}

\usepackage{xcolor}

\begin{document}

\title{Superorders and acoustic modes folding in {BiFeO$_3$/LaFeO$_3$} superlattices}

\author{R. Gu$^{1}$, R. Xu$^{2}$, F. Delodovici$^{2}$, B. Carcan$^{3}$, M. Khiari$^{3}$, G. Vaudel$^{1}$, V. Juv\'e $^{1}$, M. Weber$^{1}$, A. Poirier $^{1}$, P. Nandi $^{4}$, B. Xu $^{5}$, V. E.  Gusev$^{6}$, L. Bellaiche$^{7}$, C. Laulh\'e $^{8,9}$, N. Jaouen $^{9}$, P. Manuel$^{10}$, B. Dkhil$^{2}$, C. Paillard$^{2,7}$\footnote{ Electronic address: paillard@uark.edu}, L. Yedra$^{4}$, H. Bouyanfif$^{3}$\footnote{ Electronic address: houssny.bouyanfif@u-picardie.fr}, P. Ruello$^{1}$\footnote{ Electronic address: pascal.ruello@univ-lemans.fr}}

\affiliation{
$^{1}$Institut des Mol\'ecules et Mat\'eriaux du Mans, UMR 6283 CNRS, Le Mans Universit\'e, 72085 Le Mans,  France\\
$^{2}$Laboratoire Structures, Propri\'et\'es et  Mod\'elisation des Solides, CentraleSup\'elec, UMR CNRS 8580, Universit\'e Paris-Saclay, 91190 Gif-sur-Yvette, France
$^{3}$Laboratoire Physique de la Mati\`ere Condens\'ee, Universit\'e Picardie-Jules Vernes, Amiens, France.\\
$^{4}$ Laboratory of Electron Nanoscopies (LENS) - MIND - Department of Electronics and Biomedical Engineering  \& Institute of Nanoscience and Nanotechnology (IN2UB), University of Barcelona, Spain.\\
$^{5}$Institute of Theoretical and Applied Physics, Jiangsu Key Laboratory of Thin Films, School of Physical Science and Technology, Soochow University, Suzhou 215006, China\\
$^{6}$Laboratoire d'Acoustique de Le Mans Universit\'e, UMR CNRS 6613, Le Mans Universit\'e, 72085 Le Mans, France\\
$^{7}$ Physics Department and Institute for Nanoscience and Engineering, University of Arkansas, Fayetteville, Arkansas 72701, USA \\
$^{8}$ Universit\'e Paris Saclay, CNRS UMR8502, Laboratoire de Physique des Solides, 91405, Orsay, France
$^{9}$ Université Paris-Saclay, Synchrotron Soleil, 91190, Saint-Aubin, France.\\
$^{10}$ ISIS Neutron and Muon Facility, STFC Rutherford Appleton Lab, UK\\
}


\begin{abstract}

Superlattices are materials created by the alternating growth of two chemically different materials. The direct consequence of creating a superlattice is the folding of the Brillouin zone which gives rise to additional electronic bands and phonon modes. This has been successfully exploited to achieve new transport and optical properties in semiconductor superlattices,
for example. Here, we show that multiferroic BiFeO$_3$/LaFeO$_3$ superlattices are more than just periodic
chemical stacking. Using transmission electron microscopy, X-ray diffraction and first-principles calculations, we demonstrate the existence of a new order of FeO$_6$ octahedra, with a period along the growth direction about twice that of the chemical supercell, i.e. a superorder. The effect of this new structural order on the lattice dynamics is studied with ultrafast optical pump-probe experiments. While a mode at 1.2 THz is attributed solely to the chemical modulation of the superlattice, the existence of another 0.7 THz mode seems to be explained only by a double Brillouin zone folding in agreement with the structural description. Our work shows that multiferroic BiFeO$_3$/LaFeO$_3$ superlattices can be used to tune the spectrum of coherent THz phonons, and potentially that of magnons or electromagnons.


\end{abstract}


\maketitle

\section{Introduction}


Superlattices are artificial structures in which the periodic sequence of nanolayers defines a chemical super-order. These superlattices offer a rich playground to confine particles and quasi-particles. Through the control of the Brillouin zone-folding process it permits to tailor quantum-well electronic wavefunctions~\cite{smith} and folded-THz acoustic phonon spectrum \cite{cardona,bartels,sun,huynh,kent311,wilson,vit}. This phonon and electron band engineering in superlattices has been successfully managed with semiconductors and has led to important technological developments like the quantum cascade laser \cite{faist} and more recently the emergence of phonon lasers \cite{saser}. Beyond semiconductors, oxides play a particular role because of the plethora of functionalities already employed in devices and sensors, but also because there exists the recent capability in growing high quality oxide superlattices \cite{ramesh,das,yadav,elssaeser,bargheer,lanzillotti,carcan,mundy}. Due to the existence of multiple couplings such as (anti)ferroelectricity, (anti)ferromagnetism and ferroelasticity, the oxide superlattices exhibit more than the periodic sequence of a chemical order, unlike semiconductors. For instance, it has been revealed recently in PbTiO$_3$-SrTiO$_3$ superlattices the existence of vortices \cite{das,yadav,li}, and ferroelectric multidomains structures in oxide superlattices which, for instance, play a central role for the negative capacitance effect \cite{zubko}. 
Room temperature multiferroic BiFeO$_3$-based superlattices have been also grown only recently \cite{carcan,mundy} and reveal the existence of new polar and structural orders like anti-ferroelectric order and structural/polar orders that are incommensurate with the chemical superlattice parameter~\cite{rispens,maran1,maran2}. The short and long-range structural order, arising from the intrinsic confinement in the superlattice and the competition between interfacial stress and electrostatic boundaries, have been already studied by X-ray diffraction, electron microscopy and piezoelectric measurements \cite{ramesh,das,yadav,carcan,mundy,rispens,maran1,maran2}. 

In this work, by analysing electron and X-ray diffraction patterns, we reveal the existence of rich multiple structural orders in a BiFeO$_3$/LaFeO$_3$ (BFO/LFO) superlattice. We show that besides the known in-plane orthorhombic distortion and antiferroelectric-like order \cite{carcan,mundy}, a long-range out-of-plane super-order is revealed. We evidence a supercell parameter which approaches the double ($\sim 2\Lambda$) of the chemical superlattice parameter. The latter one ($\Lambda$) being solely defined by the thickness of the double layer BFO/LFO. On the basis of a first principle calculation, we assign this new order to a specific sequence of FeO$_6$ octahebra tilt that makes two adjacent BFO layers non-equivalent. We arrive then to a super-period BFO(1)/LFO/BFO(2)/LFO corresponding to a structural period of $\sim 2\Lambda$. We then discuss how such super-order can affect the THz acoustic phonon spectrum of the superlattice. 
For this purpose, we performed time-resolved optical pump-probe experiments. We show that with these BiFeO$_3$/LaFeO$_3$ superlattices it is indeed possible to optically generate and detect at room temperature a 1.2 THz longitudinal coherent acoustic phonon mode which is consistent with mode-folded phonons arising from the chemical order. Such a folded mode was already revealed in different oxide superlattices like in Pb(Zr,T)O$_3$-SrRuO$_3$ \cite{elssaeser,bargheer} and BaTiO$_3$-SrTiO$_3$ \cite{lanzillotti} superlattices. More interestingly, we detect a new mode centered at 0.7 THz which cannot be explained by taking into account solely the superlattice chemical order. We attribute this new mode to a secondary mode-folding process in agreement with the existence of a nearly double super-cell revealed by the structural analysis. These finding opens new avenue to tailor the spectrum of particles (electron, phonon, electromagnon) in multiferroic superlattices with a controllable mode-folding.


\section{Results and discussion}
\begin{figure}[ht!]
\centering
\includegraphics[width=0.9\textwidth]{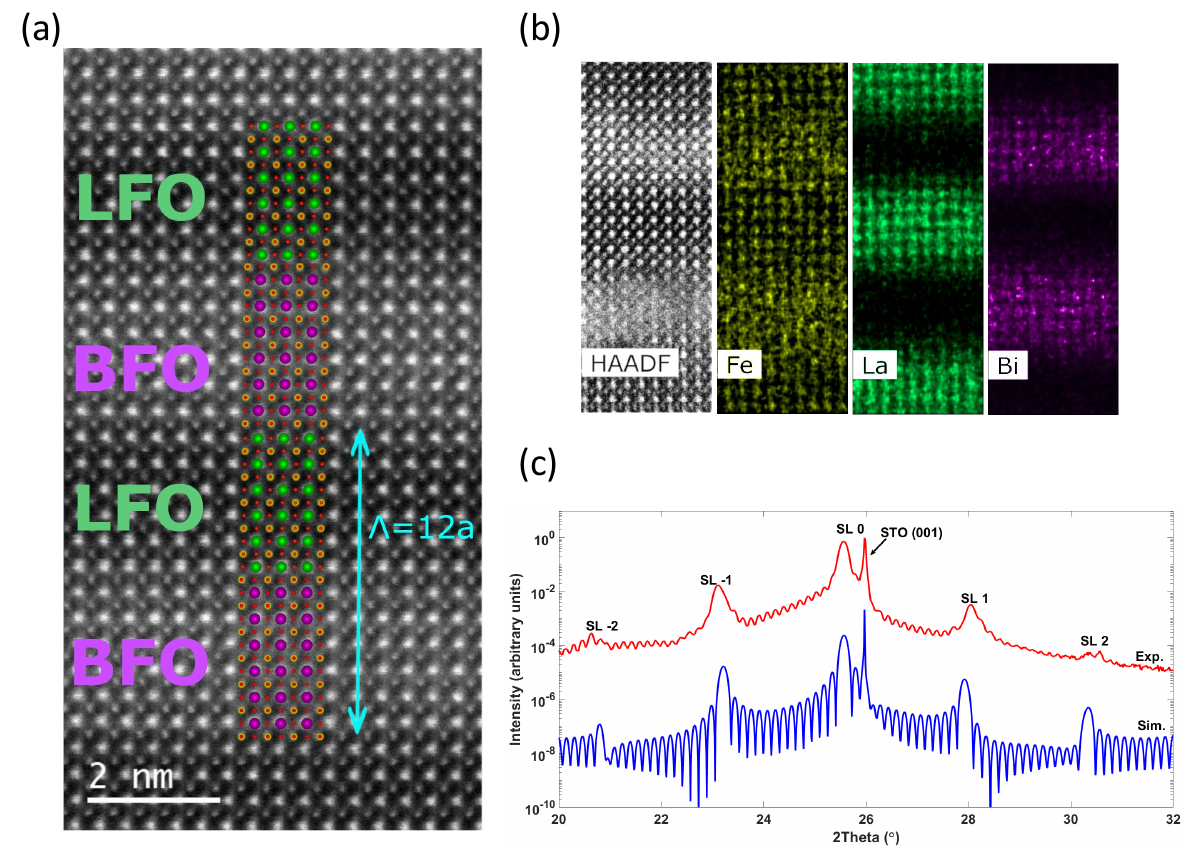}
\makeatletter\long\def\@ifdim#1#2#3{#2}\makeatother
\caption{\label{fig1}\linespread{1}\footnotesize{ (color online) (a) High Angle Annular Dark Field Scanning Transmission Electron Microscopic (HAADF-STEM) image of the BiFeO$_3$/LaFeO$_3$ superlattice. (b) spatial distribution of the chemical elements (Energy Dispersive X-ray Spectroscopy - EDS). (c) Sequence of superlattice Bragg peaks (SL $\pm$ n) measured by X-ray diffraction in the $\theta-2\theta$ geometry (red curve) corresponding to a scanning of the reciprocal space along [001]*. The blue curve is a simulation (see text). }}
\end{figure}

\subsection{Growth and structural analysis}
The BiFeO$_3$/LaFeO$_3$ superlattice has been grown with Pulsed Laser Deposition (PLD) and details can be found in Methods. On total, 15 bilayers (BFO/LFO) have been grown for a superlattice thickness of around 60~nm. The superlattice has been studied with electron and X-ray diffraction methods combined with Transmission Electron Microsocopy-TEM imaging (see Methods). A High Angle Annular Dark Field Scanning Transmission Electron Microscopy image (HAADF-STEM) is presented in Fig.~\ref{fig1}a where the chemical supercell parameter $\Lambda=d_{BiFeO_3}+d_{LaFeO_3}$ is shown with $d_{BiFeO_3}$ and $d_{LaFeO_3}$ being the corresponding layer thicknesses. It is worth noting first that the sample does not show major stacking-faults or dislocations. 
The sequence of the chemical order (Bi, Fe, La) is presented in Fig.~\ref{fig1}b where we can see a slight diffusion at the interface. Figure~\ref{fig1}c presents the X-ray diffraction pattern in $\theta/2\theta$ geometry for the BFO/LFO SL for the first order of diffraction corresponding to a scanning of the reciprocal space along $[00\xi]$*(see Methods). Intense, sharp and regularly spaced satellite peaks are observed which demonstrate a true chemical modulation of the stacking and not a solid solution. Finite-size fringes in between the satellites are also observed further highlighting the excellent structural quality of the SL. In order to get access to the out-of-plane lattice parameters of the BFO and LFO layers and the number of BFO and LFO unit cells, this pattern has been simulated with the Matlab Interactive XRDfit program \cite{simulRX} (blue curve in Fig.~\ref{fig1}c). This semi- kinematical approach allows a fast simulation by adjusting the number of unit cells and the out-of-plane lattice parameters in each layer (BFO and LFO) of the period $\Lambda$ (Note that it does not take into account interdiffusion, strain gradient and domains). BFO and LFO number of unit cells used for the simulation are 6 and 5 with a lattice parameter of 0.4~nm and 0.3925~nm, respectively. The out-of-plane lattice parameters are in agreement with an overall in-plane compressive strain for a coherent growth. Note that a structure BFO(5)/LFO(6) also gives a reasonable fit indicating that there is at the macroscopic level a fluctuation of the number of the layers with then $\Lambda \sim 11a$, where $a$ is the average pseudo-cubic lattice parameter between BFO and LFO. 
\begin{figure}[h!]
\centering
\includegraphics[width=1\textwidth]{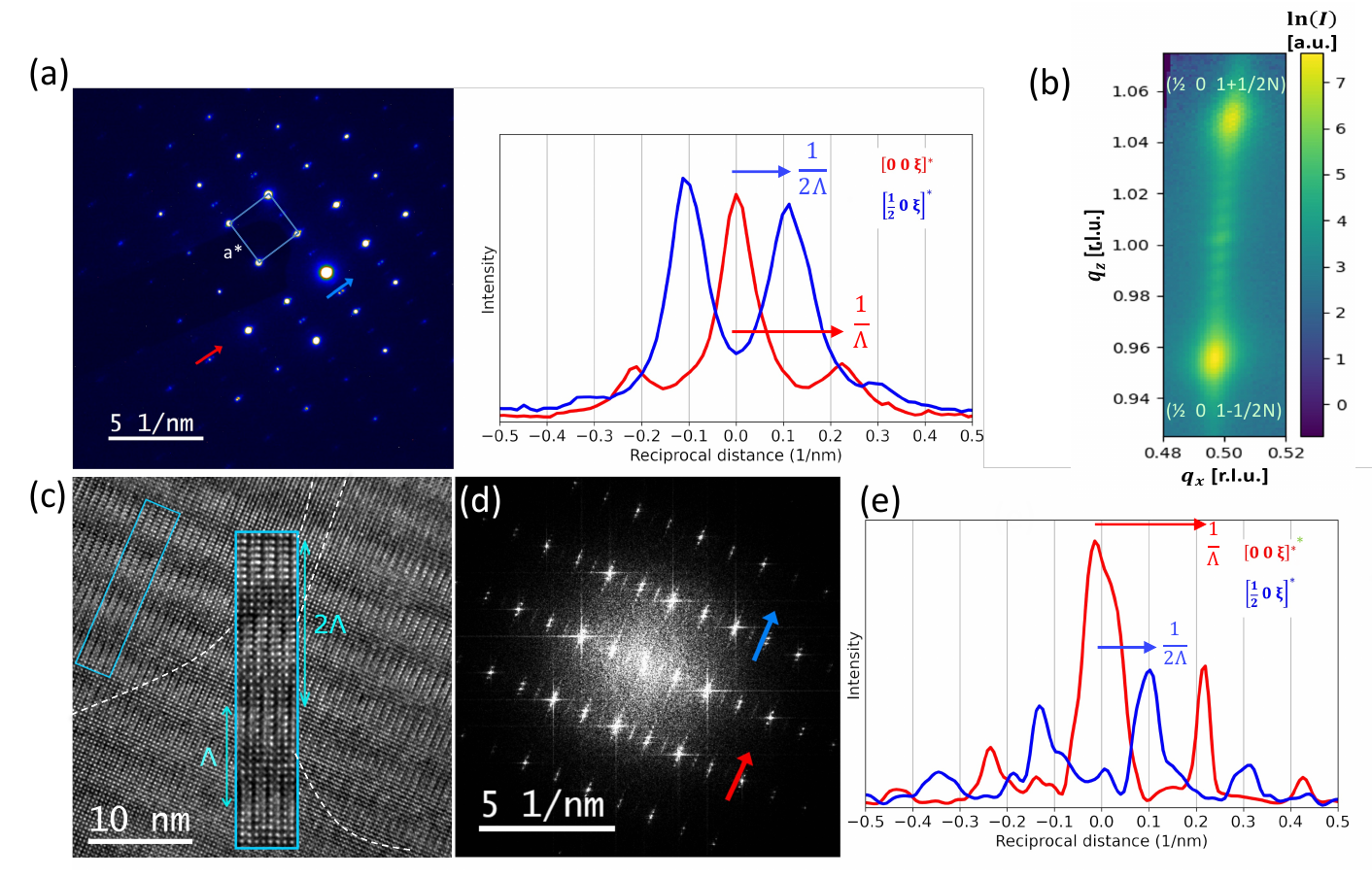}
\makeatletter\long\def\@ifdim#1#2#3{#2}\makeatother
\caption{\label{fig2}\linespread{1}\footnotesize{
 (color online) (a) (left panel) Electron diffraction measurements revealing the pseudo-cubic reciprocal lattice in the main panel ($a^{*}~=1/a$, where a is the mean lattice parameter between BiFeO$_3$ and LaFeO$_3$). When zooming along the red and blue arrows, i.e. along $[0 0 \xi]$* and $[\frac{1}{2} 0 \xi]$* reciprocal directions respectively, super-structures are revealed. 
(right panel) The corresponding intensity profiles of these super-structures reveal the chemical supercell $\Lambda~\sim Na$ (red curve) but also a double super-order $2\Lambda\sim 2Na$ (blue curve). (b) X-ray diffraction peaks along the $[\frac{1}{2} 0 \xi]$* direction confirm the existence of the super-orders ($\sim2\Lambda=2Na$). $q_x$ and $q_z$ are reduced wavevectors. The intensity modulation between each of these super-structures are assigned to Kiessig fringes. 
(c) High Resolution Transmission Electron Microscopy (HRTEM) revealing two different zones separated by the dashed line where the superlattice period can be either purely chemical ($\Lambda$) or can exhibit an additional superorder ($\sim$$2\Lambda$). (d) Fast Fourier Transform image of (c) reveals several super-structures consistent with electron diffraction image shown in (a). (e) The intensity profile around the super-structures (see green, blue and red arrows indicated in (d) for the reciprocal directions $[\frac{1}{4}0\xi]$*, $[\frac{1}{2}0\xi]$* and $[00\xi]$*, respectively) reveal the first and second order diffraction coming from the chemical order ($\Lambda$) and from the additional super-order ($\sim2\Lambda$).} }

\end{figure}

Very importantly, while Fig.~\ref{fig1}c is sensitive to the average unit cell and superlattice ($\Lambda$) parameters in the $[00\xi]$* direction only, electron and X-ray diffraction measurements in the extended reciprocal space presented in Fig.~\ref{fig2}a-b and HRTEM images shown in Fig.~\ref{fig2}c-e acquired in different regions, reveal that a more complex short and long-range atomic and/or polar arrangements are present. The electron diffraction pattern shown in Fig.~\ref{fig2}a (top left panel) shows first the in-plane unit-cell doubling with Bragg peak at $(\frac{1}{2} 0 0)$ consistently with Pnma (Pbnm) structure already reported in the literature~\cite{carcan,mundy}. The chemical supercell parameter ($\Lambda$) is evidenced along the $[00\xi]$* reciprocal space direction as revealed by the corresponding profile of electron diffraction intensity. An example of the profile of one of these Bragg peak intensities (the selected one is located by the red arrow in the top left panel) is shown by the red curve in the top right panel of Fig.~\ref{fig2}a. The observation of the super-structure ($\Lambda \sim Na $) is consistent with the chemical superstructure revealed by X-ray diffraction in Fig.~\ref{fig1}c. Importantly, when scrutinizing the Bragg peak around the position (${1+\frac{1}{2}}$ 0 $\xi$), located by the blue arrow in the top left panel of Fig.~\ref{fig2}a, two satellites are evidenced at around $(\frac{1}{2}0{\pm\frac{1}{2N}})$. The intensity profiles of these additional super-structures ($\sim2\Lambda=2Na$) are shown in Fig.~\ref{fig2}a (blue curve in the top right panel). With these measurements, a numerical estimate gives $\Lambda \sim Na=11.2a$ consistently with the value deduced from the fit of the X-ray diffraction (see Fig.~\ref{fig1}c). These satellites $(\frac{1}{2}0{\pm\frac{1}{2N}})$ demonstrate the existence of a super-order whose periodicity ($\sim 2\Lambda$) is around twice that of the chemical order. This super-order $\sim 2\Lambda$ is also observed in X-ray diffraction with the two satellites ($\frac{1}{2}$ 0 1$\pm \frac{1}{2N}$) shown in Fig.~\ref{fig2}b where a fit of these satellites position gives a characteristic length of $2\Lambda \approx21.1a$. 

High Resolution Transmission Electron Microscopy (HRTEM) images shown in Fig.~\ref{fig2}c reveal that there are statistically two regions (that we can roughly separate by a white dashed line in Fig.~\ref{fig2}c) with different orders. The central region is the so-called conventional one (no super-order) while the two others (in the left and right parts of Fig.~\ref{fig2}c) are the "unconventional ones" (with super-order $2\Lambda$). An analysis at a larger scale indicates that these two regions are statistically distributed in the superlattice and these regions have a typical size of 20~nm. 
A 2D Fast Fourier Transform Image (Fig.~\ref{fig2}d) perfectly confirms the observations made with the electron and X-ray diffraction measurements. The nearly double superlattice period ($\approx 2\Lambda$) is revealed only along the reciprocal space direction $[\frac{1}{2}0\xi]$*. Even if the intensity is smaller and more diffuse, we also detect split Bragg reflections around the position ($\frac{1}{4}$00). This indicates an anti-ferroelectric PbZrO$_3$-like structure consistently with previous reports \cite{carcan,mundy}. 
The intensity profiles shown in Fig. \ref{fig2}e confirm the two characteristic lengths $\sim \Lambda$ and $\sim 2\Lambda$. It is remarkable to note that the new super-order ($\sim 2\Lambda$) is detectable only along the $[\frac{1}{2}0\xi]$* direction while the so-called chemical super-order is evidenced along $[00\xi]$*. This means that only the regions which exhibit the associated the Pnma (Pbnm) symmetry at the unit cell level, i.e. a doubling of the unit cell in the in-plane direction, also exhibit this new out-of-plane super-order. 
When focusing in the real space (see inset of Fig.~\ref{fig2}c), we can see that in the super-ordered region, there is an intensity modulation of the electron diffraction image in the BiFeO$_3$ layer that alternates in the plane of the layer (bright-dark Bi columns). Very interestingly, this sequence also exists in the adjacent BFO layer but is in anti-phase as perfectly revealed in the light blue rectangle in the inset of Fig.~\ref{fig2}c. This intensity modulation in the real space is consistent with our description of a double super-cell parameter. 
\begin{figure}[h!]
\centering
\includegraphics[width=0.9\textwidth]{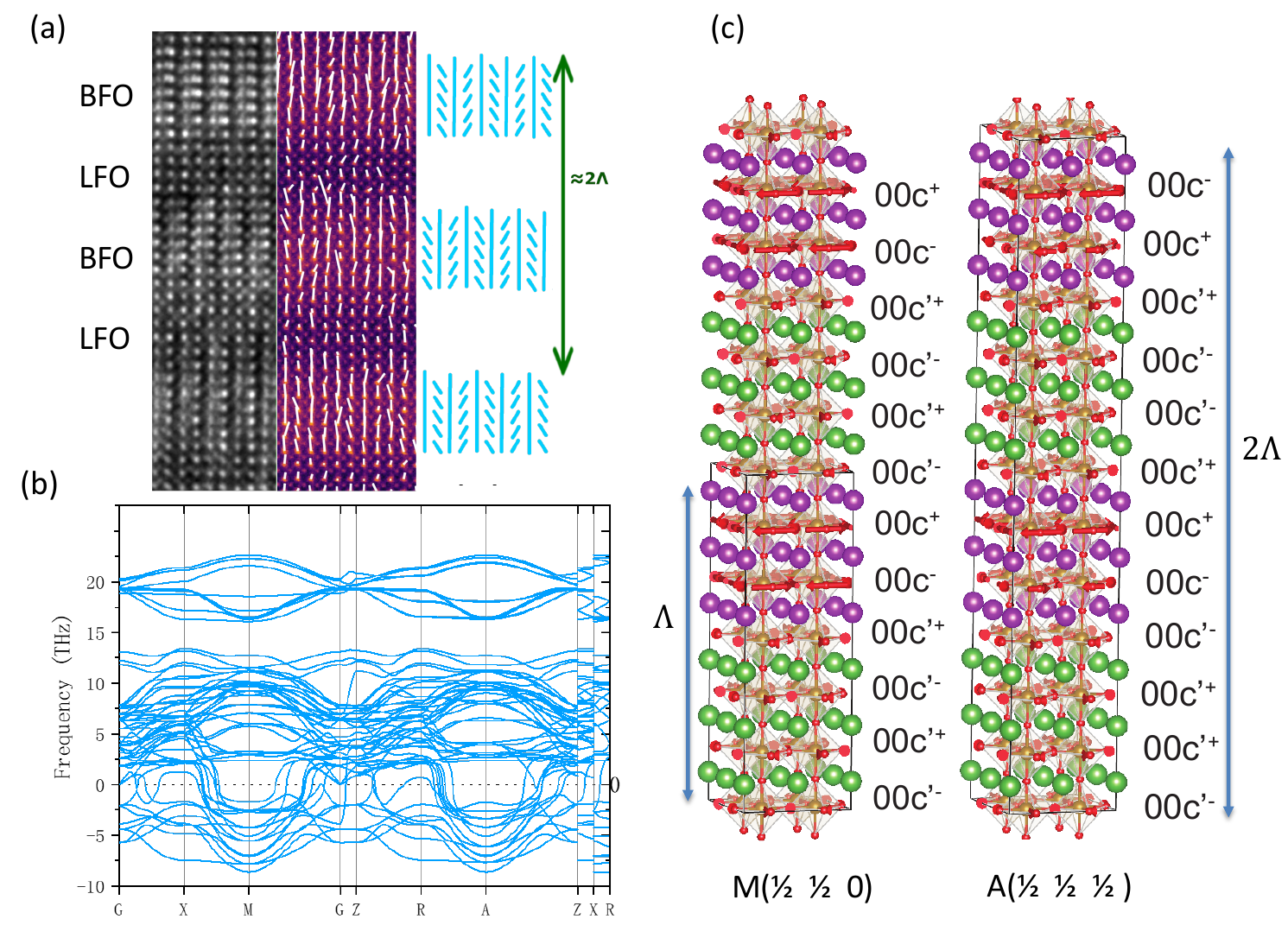}
\makeatletter\long\def\@ifdim#1#2#3{#2}\makeatother
\caption{\label{fig3}\linespread{1}\footnotesize{ (color online) (a) HRTEM image and A-site displacements in the unconventional region extracted from a HAADF-STEM image performed at 300K. The sketch on the right panel depict the non-equivalence of the Bi displacement (light blue lines) between two adjacent BFO layers. (c) Phonon dispersion in the high-symmetry, "cubic-like" phase of the BFO(3)/LFO(3) superlattice. Imaginary frequencies representing atomic displacement that would lower the energy are depicted negatively (soft modes). (d) Representation of the soft mode for the M and A Brillouin zone direction depicting the evolution of the cubic structure towards a conventional (M) and superordre (A) structure. The DFT calculation is realized with nanolayers of BFO (Bi atomes in purple) and LFO (La atoms in green) having 3 unit cells.}}
\end{figure}
A closer look on displacement of the Bismuth atoms (Fig.~\ref{fig3}a) in the "unconventionnal region" has also been made in the BFO layers with High-Angle Annular Dark-Field images (HAADF-STEM). In the BFO layer, there is a well structured organization of this displacement. Some Bismuth atoms exhibit a vertical displacement and in between some of them rather exhibit an oblique displacement, that is either left or right oriented. It is worth mentioning that the apparent displacement of the Bismuth atoms (vertical or oblique) is a consequence of a projection of displacement of several Bismuth atoms in the atomic column between the top and bottom surfaces of the TEM lamella. This leads to an elliptical distortion of the projected Bismuth atomic column as previously discussed~\cite{mundy}. The important new element we report is the anti-phase organization of this displacement for two adjacent BFO layers that explains the existence of a supercell parameter $\sim 2\Lambda$. This effect can be observed in the work of Mundy et al~\cite{mundy} but has not been discussed so far. 

\subsection{Modeling of the superlattice structure}
In order to further delve into the origin of the doubling of the supercell, we calculated the phonon dispersion in high-symmetry, "cubic-like" BFO$_n$/LFO$_n$ ($n=1,2,3$) superlattices using Density Functional Perturbation Theory (See also Supplementary Note 2). Such approach yields imaginary phonon modes (represented as negative in figure~\ref{fig3}b), which are hallmarks of atomistic distortions that develop to form the real ground state order of these superlattices. Interestingly, the most unstable phonon frequency are found at the point $A=(1/2,1/2,1/4n)$ and $M=(1/2,1/2,0)$ in pseudocubic reciprocal lattice notation. In particular, for the largest superlattices investigated ($n=3$), the $A$ and $M$ instabilities almost have the same imaginary frequency. If such features were to persist at $n=6$, it would explain why we observe two regions, corresponding to a separate condensation of the two modes. In such a case, the "conventional" region could be attributed primarily to the condensation of the $M$ phonon mode (region without super-orders), while the unconventional region with $1/24$ modulation would be very well explained as resulting from the condensation of the $A$ phonon instability (region with super-orders). The associated soft-modes with the corresponding sequence of FeO$_6$ octahedra tilts (using Glazer's notation~\cite{Glazer1972}) in the out-of-plane direction are shown in Fig.~\ref{fig3}c. Similar octahedra tilts sequences were theoretically predicted in bulk materials as a possible new orders in perovskite oxides~\cite{prosandeev2013,bellaiche2013} and we very likely provide the first experimental evidence of existence of these orders in BiFeO$_3$/LaFeO$_3$ superlattices. 



\subsection{Coherent phonon dynamics}
\begin{figure}[h!]
\centering
\includegraphics[width=0.9\textwidth]{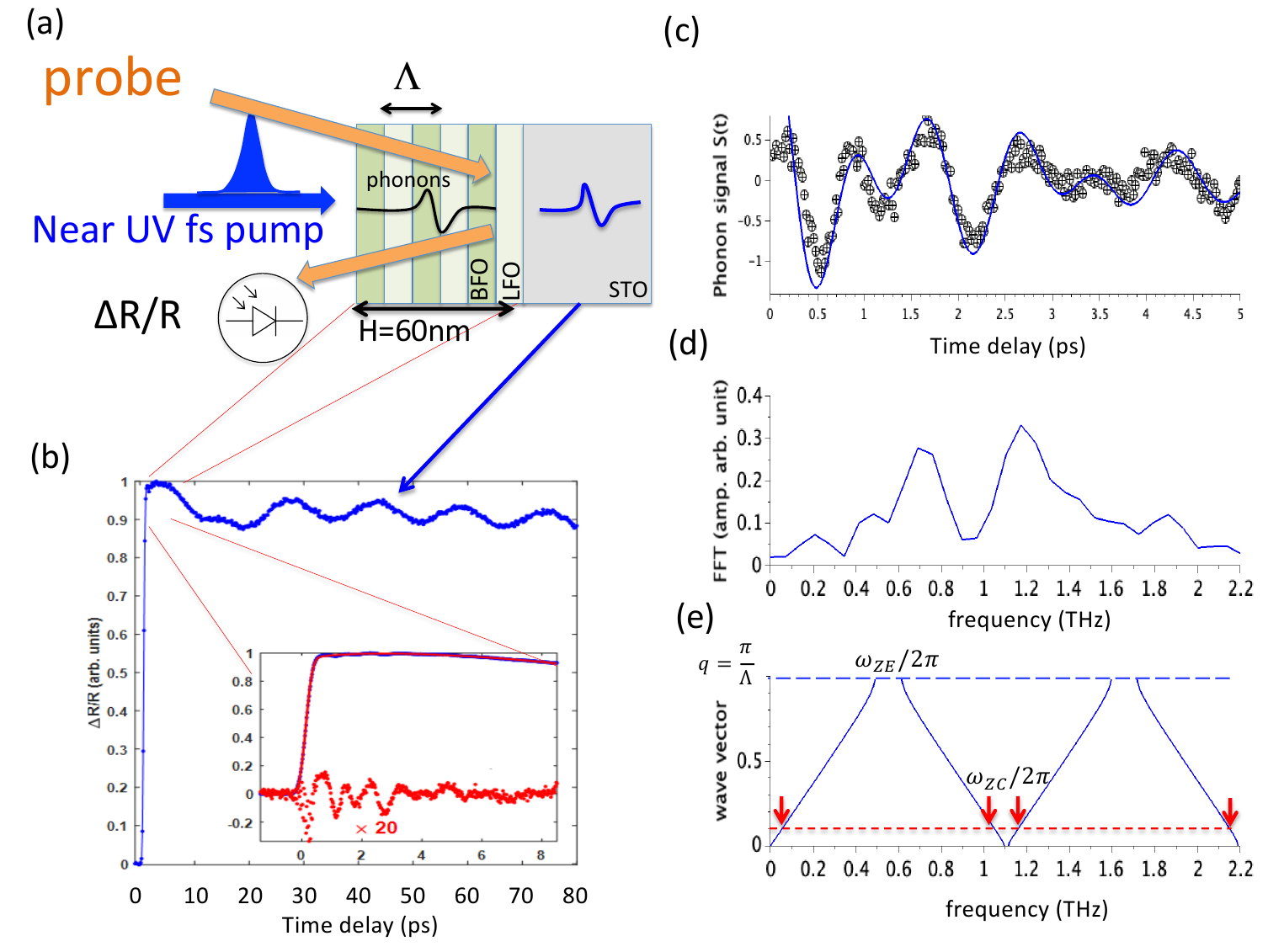}
\makeatletter\long\def\@ifdim#1#2#3{#2}\makeatother
\caption{\label{fig4}\linespread{1}\footnotesize{ (color online) (a) Principle of the pump-probe scheme (b) Typical transient optical reflectivity signal revealing at long time scale the Brillouin component in the SrTiO$_3$ substrate, while the short time scale (see inset) reveals the THz acoustic phonons generated and detected in the superlattice. (c) Numerical adjustment of the THz acoustic phonons signal with two sinusoidal functions (see text for details). (d) Fast Fourier Transform of the signal shown in (c). (e) acoustic phonon band dispersion curves of a superlattice having a supercell parameter (chemical order) of $\Lambda$. }}
\end{figure}

These rich structural super-orders are a perfect demonstration that in such a BFO-LFO superlattice, additionally to the chemical order i.e. the sequence of the BFO and LFO layer, there is an additional long-range structural order ($\approx 2\Lambda$). Such nearly double supercell might have an important consequence on the spectrum of THz coherent acoustic phonons as discussed in the following. To support it, we have investigated the generation and detection of coherent acoustic phonon following an ultrafast optical pump-probe method (Fig.~\ref{fig4}a). This technique has been previously employed to study the coherent acoustic phonons in bulk BiFeO$_3$ for instance~\cite{lejman2014,lejman2016,lejman2019,juve2020}. The pump beam, with photon energy above BFO and LFO electronic band gaps, is absorbed and generate acoustic phonons propagating only along $[001]$ direction (see Methods). The time-delayed probe beam, with photon energy below these electronic band gaps, is back-scattered by the propagating photoexcited phonons (see Methods). The analysis of the transient optical reflectivity permits to reveal these coherent acoustic phonons through characteristic Brillouin oscillations (time-domain Brillouin spectroscopy). Since the detection of the coherent acoustic phonons is realized in the reflection geometry (back-scattering) and with a normal incidence, due to the momentum conservation, the coherent acoustic phonons that are detected in the Brillouin process are those at a wave vector two times that of the probe beam, with $q=2k_{probe}$. Since the probe wavelength $\lambda$ is much larger that the lattice $a$ and superlattice $\Lambda$ parameters, the detection takes place very close to the Brillouin zone center (ZC), i.e. with $q\sim0$. 

The typical signal shown in Fig.~\ref{fig4}b is composed of a fast rise of the transient optical reflectivity signal corresponding to the electronic excitation of the BFO-LFO system. Then a plateau-like response is observed (long-lifetime of photoexcited carriers and thermal effect) on which oscillatory signals are revealed due to the Brillouin process~\cite{lejman2014,lejman2016,lejman2019,juve2020}. The first detected period over a long time delay up to 80~ps is around 12\,ps corresponding to a Brillouin frequency of 60~GHz. It is important to have in mind that with a sound velocity in BFO of around 5000 m.s$^{-1}$~\cite{lejman2014,lejman2016,lejman2019,juve2020} and quite similar for LFO \cite{mahdi}, after typically 10\,ps the photogenerated acoustic phonon has left the superlattice. This means that the Brillouin signal at 60~GHz come from the detection of acoustic phonon in STO. That frequency is perfectly consistent with the theoretical value $f_{STO}={2n_{STO}V_{STO}}/{\lambda}$=63GHz , where $n_{STO}=2.4$ and $v_{STO} \approx 8300$ m.s$^{-1}$ are the refractive index and the longitudinal sound velocity in STO~\cite{maerten}. More interestingly, within the first tens of picoseconds (inset of Fig.~\ref{fig4}b), we reveal oscillatory components having much higher frequencies with periods of a couple of picoseconds. 
A zoom on this acoustic phonon signal is given in Fig.~\ref{fig4}c for a probe wavelength of 587nm. The Fast Fourier Transform (FFT) given in Fig.~\ref{fig4}d reveals mainly two modes with $f_1=$0.7 THz and $f_2=$1.2 THz. A simple simulation of this signal ($S(t)$) reproduces quite well the experimental observations as shown in Fig.~\ref{fig4}c with $S(t)=(A_1cos(w_1t)+A_2cos(w_2t))e^{-t/2ps}$, considering $w_1=2\pi f_1$ and $w_2=2\pi f_2$. The damping time of 2~ps has been chosen consistently with the lifetime observed experimentally. The amplitude has been taken as $A_2=1.15 A_1$ and the phase is zero for both modes. As discussed in the following, the detected THz acoustic phonons come actually from the BFO-LFO superlattice. Detecting THz acoustic phonon in superlattices with supercell parameter of few nanometers is indeed expected due to the mode folding mechanism several times discussed in the literature for semiconductors and oxides \cite{cardona,bartels,sun,huynh,kent311,wilson,vit,elssaeser,bargheer,lanzillotti}. It is however the first ever reported coherent THz acoustic phonon generated and detected in BiFeO$_3$-based superlattice. With a set of elastic properties of BFO and LFO (either calculated or extracted from the bulk system, See Supplementary Note 2), it is possible to analytically calculate the phonon dispersion curve $\omega=f(q)$ for longitudinal acoustic phonon propagating along the out-of-plane direction and for the superlattice period $\Lambda=d_{BFO}+d_{LFO}=12a$. The dispersion curve is shown in Fig.~\ref{fig4}e where we see the mode folding of the LA mode in this BFO-LFO superlattice with the Brillouin zone edge vector $\pi/\Lambda$. In that case, the region indicated by the red arrows (Fig.~\ref{fig4}e) are the ones where some coherent acoustic phonon are expected to be detected according to the aforementioned selection rule ($q~=~2k_{probe}~\approx0$). At the very low frequency ($<$100 GHz) a Brillouin mode is expected to be detected in the SL (similarly to the one observed in STO) but is not since its characteristic period of oscillation (typically the one observed for STO) is longer than the time of flight of the acoustic phonons in the SL. More importantly, there is a second region ($\omega_{ZC}$) where modes can be detected and values of around 1.1-1.2 THz are expected. These frequency are actually consistent with the detected mode $f_2$ with $\omega_{ZC}/2\pi \sim f_2$. 
Such $\omega_{ZC}$ zone centered mode coming from the mode folding governed by the periodic chemical composition BFO-LFO ($\Lambda$) is the typical one that is conventionally observed in the coherent phonon spectra of many semiconductor \cite{cardona,bartels,sun,huynh,kent311,wilson,vit} and ferroelectric \cite{elssaeser,bargheer,lanzillotti} superlattices. Very importantly, we show that the mode $f_1$=0.75 THz cannot be explained by the conventional mode folding driven by the chemical super-order. In the theoretical calculation, a mode in the region 0.5-0.7 THz is predicted, but it is a zone edge mode ($\omega_{ZE}$), thus not detected according to the optical detection selection rule. We discuss in the following the origin of this mode.

It is first worth mentioning that magnon mode at 0.7 THz has already been reported in BiFeO$_3$ nanostructures~\cite{sandocyclo} arising from a magnon mode folding caused by the magnetic cycloid. Such a mode can be excluded in our case since we have a confined and strained multiferroic nanostructure where the cycloid is no more present as discussed in the literature~\cite{sandocyclo}. We have confirmed the absence of the cycloid with polarized neutron diffraction in a BFO(12)/LFO(12) superlattice (see Supplementary Note 3). 

The origin of this mode at 0.7 THz might have another origin. Different hypothesis can be discussed. First of all, one could envision also a shear acoustic mode folding. Considering the shear velocity of around 3000~m.s$^{-1}$~\cite{lejman2014,juve2020,ruello2012,gu2023}, one can expect a zone center mode at around 0.7~THz (calculation not shown). But, generating a shear acoustic phonon requires that the system has the proper symmetry, i.e. that an in-plane symmetry breaking exists (i.e. absence of a symmetry plane perpendicular to the surface). This condition is fulfilled in the bulk single-domain BiFeO$_3$, with rhombohedral symmetry ($3m$)~\cite{lejman2014,juve2020,ruello2012} or in single domain monoclinic BiFeO$_3$ thin film~\cite{gu2023}. This in-plane symmetry breaking is usually well correlated with optical birefringence in both aformentioned cases as a witness of crystal anisotropy~\cite{lejman2014,juve2020,gu2023}. Such optical birefringence is absent in the BFO/LFO superlattice (see Supplementary Note 4). This is an indication, that at the level of an optical beam focus, the BFO/LFO superlattice is isotropic. This is consistent with the microscopic organisation revealed by electron microscopy where we see the coexistence of so-called conventional and non-conventional regions (see discussion in Structural analysis Section). We do not have a single domain ferroelectric/multiferroic structure in our superlattice. Even if at the level of the unit cell, light can generate shear acoustic waves, the multidomain state might lead to an average effect at the macroscopic level. So we think then that we can exclude that the mode at 0.7 THz comes from the generation and detection of coherent transverse mode. 
Secondly, we have also to consider that the mode at 0.7~THz could be a localized surface mode as revealed in Mo/Si superlattice for example~\cite{perrin2009}. Such localized mode usually appears in superlattices having a wide forbidden phonon band gap due to a very large acoustic impedance mismatch between the layers and terminated with the lower acoustic impedance layer~\cite{perrin2009}. In our case, the top layer LaFeO$_3$ has a lower acoustic impedance than BiFeO$_3$ (see Supplementary Note 2) but the forbidden band gap is much smaller than the one observed in Mo/Si layer and our mode at 0.7 THz does not fall into the phonon bandgap. As a consequence, at the moment, it is difficult to assign the mode 0.7 THz to a localized mode. 
Finally, we think that a plausible interpretation could rely on the new super-ordered we reveal in this superlattice. With an out-of-plane period that is around $2\Lambda$ (see Fig.~\ref{fig2}), such a double superlattice period should lead to a second mode folding. In that case, this will transfer the ZE mode at the ZC position. Consequently, the so-called ZE mode would be optically detectable. When looking at the theoretical (0.5-0.7 THz) and experimental (0.7 THz) values, we see that the agreement is not perfect in our case but could be explained by the possible departure of the elastic parameters of bulk materials used in this calculation. We have also to underline that the value of the double super-period $2\Lambda$ is not perfectly an integer $a$, since we extract a value of 21.1a in X-ray diffraction (Fig.~\ref{fig2}b). This indicates that we cannot completely exclude incommensurate superorders \cite{rispens,maran1,maran2} that would contribute to a departure from a perfect double mode folding. Even if this disorder might exist, the long range order ($2\Lambda$) remains strong to be clearly revealed by electron and X-ray diffraction. Such double mode folding open very interesting perspectives for engineering the spectrum of phonons and other particles in multiferroic superlattices in the future. 

\section{Summary}

To summarize, in this work we have scrutinized the structural properties and coherent acoustic phonon dynamics in BiFeO$_3$-LaFeO$_3$ based superlattices with a combination of electron and X-ray diffraction, optical time-resolved pump-probe method and modeling. Besides the Pnma distorsion and the anti-ferroelectric orders visible at the unit cell, we show the existence of a long-range interaction whose characteristic length is at least twice the chemical supercell parameter (made of a bilayer BFO-LFO). Simulations based on DFT calculations support the existence of non-equivalent neighbor BFO layers that leads to the doubling of the superlattice parameter. Such doubling is due to the fact that both the polar order and the sequence of the oxygen octahedra tilts in the out-of-plane direction do not follow the chemical order. Rather, the pattern of the octahedra tilts scales with the double of the chemical supercell. We have evaluated the impact of this new crystallographic order on the lattice dynamics with time-resolved pump-probe experiments. We demonstrate that in this superlattice it is possible to generate coherent THz acoustic phonons (1.2 THz, 0.7 THz), never reported so far. While the mode at 1.2 THz is correlated to the chemical supercell, the detected mode at 0.7 THz seems to be connected to the newly created FeO$_6$ octahedra superorder. These results demonstrate that multiferroic superlattices are more than a stacking of chemically different layers. These findings enable to envision in the future to tailor the electron, phonon, magnon or electromagnon spectra in the THz range.\\



\textbf{Experimental Section}\\
\textbf{Sample preparation}\\
BFO/LFO superlattices (SLs) were grown directly on Niobium doped (0.5 at$\%$) (001)-oriented SrTiO$_3$ substrates by Pulsed Laser Deposition (PLD) using a KrF excimer laser (wavelength 248\,nm). A stoichiometric homemade ceramic target of LaFeO$_3$ was used whereas an extra and usual amount of 10$\%$ Bi was introduced to the homemade BiFeO$_3$ target to compensate bismuth volatility. The SLs were grown at 3Hz pulse frequency, a fluence of 1.8\,J/cm$^2$, a temperature of 775$^{\circ}$C and oxygen pressure of 0.2\,mbar. These optimized growth conditions were obtained from single BFO and LFO thin films synthesis. A low deposition rate was obtained for both compounds enabling a good control on the amount of deposited matter (53 laser shots to complete one unit cell for BFO and 73 laser shots for LFO).

The multiferroic BiFeO$_3$/LaFeO$_3$ superlattice have been grown on Niobium doped (0.5at$\%$) SrTiO3 substrate by pulsed laser deposition (excimer KrF : 248nm) at 3Hz pulse frequency, a fluence of 1.8J/cm$^2$, 725$^{\circ}C$ and oxygen pressure of 0.2mbar. Details of the growth parameters can be found in Ref. \cite{carcan}. Contrary to the previous work on BFO-LFO superlattice \cite{carcan}, no buffer layer was used. In complement to the Fig. \ref{fig1}c, the coherent growth is also verified through reciprocal space mapping around the (103) family of planes (not shown).

\textbf{Electron microscopy}\\
TEM cross-sectional lamella preparation was performed in a FEI Helios 660 FIB-SEM. Observation was carried out in two microscopes: High Resolution Scanning Transmission Electron Microscopy (HRSTEM) and Energy Dispersive X-ray Spectroscopy (EDS) were obtained in a FEI Titan3 60-300, operated at 300kV, while Electron Diffraction (ED) and High Resolution Transmission Electron Microscopy (HRTEM) were obtained in a JEOL 2010F, operating at 200kV. The analysis of the atomic lattice (atomic column position and shape) was carried out using Atomap software. 

\textbf{X-ray diffraction}\\
Synchrotron X-ray diffraction data were collected at the CRISTAL beamline of SOLEIL Synchrotron (Saint Aubin, France). The X-ray photon energy was set to 7.064 keV using a Si(111) double crystal monochromator, harmonic rejection being ensured by two reflections on Si mirrors. The diffraction signal was recorded with the 2D detector ImXPAD-S540 placed at a distance of 818.5 mm from sample position. The $\theta$-2$\theta$¸ scan presented in Figure \ref{fig1}c was obtained by summing diffracted intensities over a region of interest of 20 x 20 pixels (2.6 mm $\times$ 2.6 mm). The reciprocal space map shown in Figure \ref{fig2}b was calculated using the BINoculars software, from an angular scan performed in grazing incidence geometry. The incidence angle of the X-ray beam was fixed to 1$^{\circ}$ with respect to the (001) surface plane, resulting in an X-ray footprint of 1.7 mm $\times$ 0.5 mm on the sample. The diffraction condition was tuned by rotating the sample about its surface normal, with the detector set to the relevant elevation and azimuthal angles.

\textbf{Modelling}\\
DFT calculation : We performed Density Functional Theory (DFT) calculations using the plane-wave VASP software package. We used the PBEsol exchange and correlation functional and $+U$ correction. Details of the calculations will be found in the Supplementary Note 1.

\textbf{Time-resolved pump-probe experiment}\\
A Ti:sapphire laser delivering laser pulses with 140fs duration at 800 nm is separated in two arms. A first one is doubled in energy in a Baryum BetaBorate crystal (BBO) according to the second harmonic generation process (SHG). This near-UV beam is used as a pump beam and has photon energy (3.1eV) larger than both the BFO \cite{optbfo} and LFO materials \cite{optlfo} bandgaps. Consequently the light is absorbed in the superlattice and not in STO having a larger band gap (3.3eV). The second 800 nm laser beam synchronously pump an optical parametric oscillator delivering visible light pulses at 580 nm (below band gap detection), to detect the phonon in the entire structure (superlattice and the STO substrate). Since the probe beam is below the band gap of BiFeO$_3$ and LaFeO$_3$, the detection mainly reveals the so-called Brillouin mode and more details about time-resolved Brillouin will be found in previous works \cite{lejman2014,lejman2016,lejman2019,juve2020}. A sketch of the experiment is shown in Fig. \ref{fig4}a. Since the pump and probe beam foot print are much larger than the typical propagation distance of detected phonon ($<$ 1$\mu$m), the generation of strain has a piston-like shape, i.e. we have a 1D geometry and the phonon propagate perpendicularly to the irradiated surface (i.e. along the $[001]$ direction). It is worth to note that Raman scattering was not possible due to the detrimental contribution of the STO substrate and that limitation is prevented with coherent phonon dynamics investigated with pump-probe methods since coherent phonon are generated in the SL only. More details on the experimental methods can be found in our previous works \cite{lejman2014,lejman2016,lejman2019,juve2020}. 
\newpage
\section{Supplementary Material}
\section{Supplementary Note 1: DFT calculation}

\begin{figure}[t!]
\centerline{\includegraphics[width=18cm]{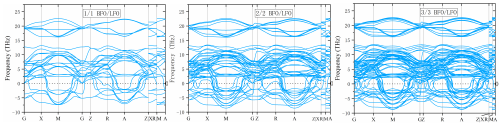}}
\caption{\label{figCharles3} (color online) Phonon dispersion curves of (left) 1/1  (center) 2/2 and (right) 3/3 cubic-like BFO/LFO SLs.}
\end{figure}

	We performed Density Functional Theory (DFT) calculations using the plane-wave VASP software package.  We used the PBEsol exchange and correlation functional and $+U$ correction. The Hubbard $U$ is 4 eV on Fe-$d$ states and 2 eV on La-$f$ states. The exchange parameter $J$ was set to $0$. The calculation was performed with a $9\times9\times9$ Monkhorst-Pack k-mesh and a 500 eV plane-wave cutoff. The magnetic order was set to G-type AFM collinear spin in the BFO/LFO superlattices. We started from the high symmetry,  cubic-like structure to see the instabilities in the phonon dispersion at 0 K. We built 1/1, 2/2, 3/3 BFO/LFO superlattices, fixing all atoms in their ideal cubic position and only relaxing along the c lattice direction. All configurations are obtained with $P4/mmm$ symmetry.  The phonon calculations are based on the density functional perturbation theory (DFPT) with the help of PHONOPY code.  The force constant matrix was calculated with a $2\times2\times2$ supercell which contains 80, 160, and 240 atoms respectively. In Figure~\ref{figCharles3}, we show the phonon spectra for the cubic-like phase of 1/1, 2/2 and 3/3 BFO/LFO SLs. Major instabilities can be observed at the $M$ and $A$ point. These correspond, respectively, to $(1/2,1/2,0)$ and $(1/2,1/2,1/2)$ reciprocal wavevectors of the BFO\textsubscript{n}/LFO\textsubscript{n} SLs.  Since the SL is made of $2n$ perosvkite layers in the third direction, the $A$ point actually corresponds to a $4n$ modulation in terms of the perovskite unit cell, \textit{i.e.} $A = (1/2,1/2,1/4n)$ in the perovskite pseudocubic notation. In particular,  this $A$ corresponds to tilt with a modulation that doubles the chemical lattice constant.

\newpage
\section{Supplementary Note 2: analytical calculation of phonon dispersion curve}

We have considered the standard equation to calculate the phonon dispersion curve $\omega=f(q)$\cite{cardona,bartels,sun,winter,huynh,kent311,wilson,vit,elssaeser,bargheer,lanzillotti}. For a superlattice composed of an alternating BFO and LFO layer, and for phonon propagating along the out-plane direction, i.e. perpendicular to the layers, the relation dispersion is : 
\begin{eqnarray}
\label{wk}
 cos(q\Lambda)&=&cos(\frac{\omega d_{BFO}}{V_{BFO}})cos(\frac{\omega d_{LFO}}{V_{LFO}})-\frac{1}{2}(\frac{Z_{BFO}}{Z_{LFO}}+\frac{Z_{LFO}}{Z_{BFO}})sin(\frac{\omega d_{BFO}}{V_{BFO}})sin(\frac{\omega d_{LFO}}{V_{LFO}})\nonumber
\end{eqnarray}

$\omega$ and $q$ are the phonon pulsation and the wave vector, respectively. The superlattice period is $\Lambda=d_{BFO}+d_{LFO}$, and  $Z_{BFO}=\rho_{BFO}V_{BFO}$ and $Z_{LFO}=\rho_{LFO}V_{LFO}$ are the acoustic impedance of each layer with $\rho_{BFO}$, $\rho_{LFO}$,  $V_{BFO}$ and $V_{LFO}$ the mass density and the longitudinal sound velocity of BFO and LFO, respectively. 

This equation can only be solved by numerical calculation. To plot the dispersion curve shown in Fig. 4d in the main manuscript, we have taken the following values that permit to reproduce the zone center (ZC) mode detected at $\sim$1.15 THz  : $\rho_{BFO}$=8220~kg.m$^{-3}$, $\rho_{LFO}$=6400~kg.m$^{-3}$,  $V_{BFO}$=5000~m.s$^{-1}$~\cite{shang,ruello2012}and $V_{LFO}$=4500~m.s$^{-1}$~\cite{mahdi}. We have considered $d_{BFO}=d_{LFO}$=0.39$\times$5.5~nm. 


\section{Supplementary Note 3: polarized neutron diffraction}
\begin{figure}[t!]
\centerline{\includegraphics[width=18cm]{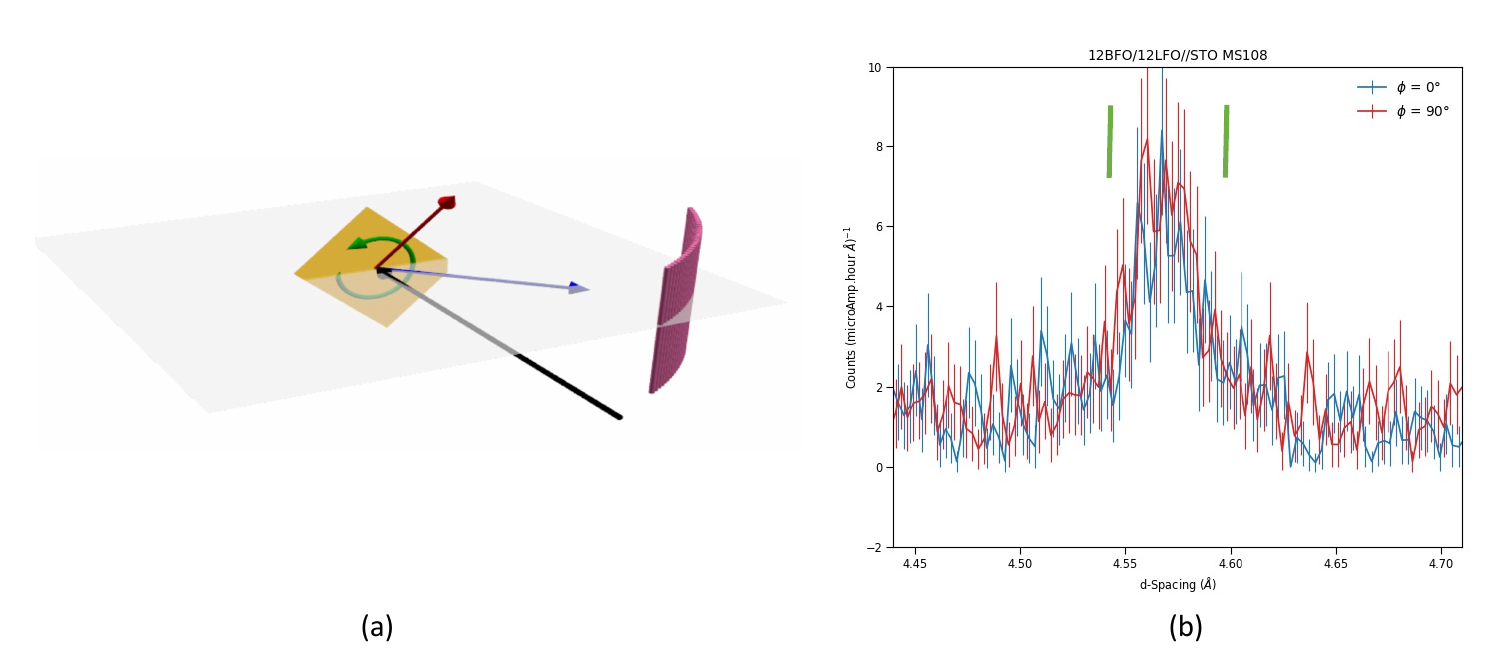}}
\caption{\label{figNeutron} (color online) (a) geometry of the measurement with a rotation of the sample by 90$^{\circ}$ to explore all the reciprocal direction where the cycloid wave vector could exist. (b) ($\frac{1}{2} \frac{1}{2} \frac{1}{2}$)  Neutron Bragg peak with the absence of the splitting (green tick) witnessing the absence of a cycloid. }
\end{figure}
Neutron diffraction data were collected on the long wavelength time-of-flight diffractometer WISH \cite{chapon} located at the second target station of the ISIS Pulsed Neutron and Muon Source, United Kingdom. By nature, neutron diffraction allows the average structure of the whole superlattice (and the substrate) to be measured. The experimental setup used was similar to our previous work on BiFeO$_3$ thin films \cite{chauleau} (Fig. \ref{figNeutron}a) but with higher spatial instrumental resolution (divergence 0.17$^{\circ}$ instead of 0.25$^{\circ}$) to allow better detection of possible peak splitting. The superlattice under investigation has a BFO and LFO layers twice thicker than the superlattice discussed in the main text. This permits to enhance the neutron signal.  Despite the fact that the structure appears antiferromagnetic, implying BiFeO3 should therefore not host a cycloid, we collected neutron diffraction data for two different rotations. Both datasets show no peak splitting either spatially or in d-spacing (along the time of flight direction) as shown in Fig. \ref{figNeutron}b  but instead reveal a single sharp peak for each orientation at the exact same d-spacing. The results are consistent with a simple G-type antiferromagnetic structure with a propagation vector [$\frac{1}{2} \frac{1}{2} \frac{1}{2}$] in cubic setting. This is in contrast to our previous measurements on thin films of BiFeO$_3$ grown on orthorhombic DyScO$_3$ substrate where clear peak splitting was observed spatially and in d-spacing \cite{chauleau}, corresponding to the tick green marks on Fig. \ref{figNeutron}b, from which the period of the modulation could be easily extracted. The fact that only one peak is observed in both datasets also reinforces the point that there is no polar direction in the system. These results show and confirm that the strong confinement prevent the formation of the cycloid and therefore magnon can no more be folded. In particular, the magnon mode at 0.7 THz cannot exist. This constraint will be all the more strong in a superlattice having even thinner BFO and LFO layers.

\section{Supplementary Note 4: optical birefringence}
\begin{figure}[t!].
\centerline{\includegraphics[width=14cm]{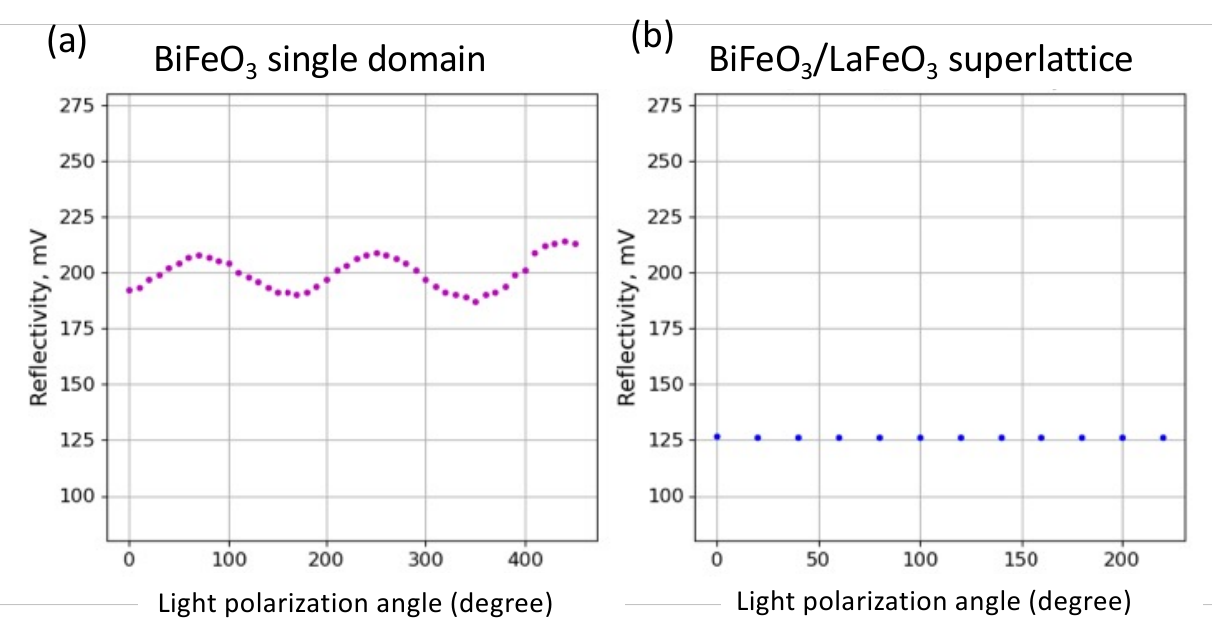}}
\caption{\label{figbi} (color online) (a) Optical reflectivity measurement performed with a single ferroelectric domain BiFeO$_3$ film~\cite{gu2023}. (b) Optical reflectivity of the BFO/LFO superlattice as a function of the probe polarization angle.}
\end{figure}

The optical reflectivity measurement has been performed with a photon energy of 3~eV, i.e. above the band gap of BFO and LFO. The light polarization was controlled with a half-plate (light polarization angle). The single domain BiFeO$_3$ thin film with a thickness of 180~nm clearly shows an optical birefringence (Fig.~\ref{figbi}a) as previously discussed \cite{gu2023}. On the opposite the optical reflectivity measured on the BFO/LFO superlattice does not show any detectable birefringence (Fig.~\ref{figbi}b). 
\newpage
%
%
%
%
\textbf{Acknowledgements} \par 
R. G, M. K., V. J, G. V. M. W, C. P., B. D. H. B, A. P. and P. R. acknowledge the French National Research Agency (ANR) under the THz-MUFINS project (Grant No. ANR-21-CE42-0030). We wish to warmly thank Fr\'ed\'eric Picca (SOLEIL Synchrotron) for adapting BINoculars to the analysis of diffraction data from CRISTAL beamline. We acknowledge SOLEIL for provision of synchrotron radiation at CRISTAL beamline (proposal number 99190273). H. B. is grateful to C. Lichtensteiger for help on the XRD simulation. C. P. and L. B. would like to acknowledge the U.S. Department of Defense under the DEPSCoR program (Award No. FA9550-23-1-0500) and the Vannevar Bush
Faculty Fellowship (VBFF, Grant No. N00014-2021-2834). L. Y. and P. N. acknowledge support from AGAUR (Spain) through 2021 FI-B00157 grant. L.Y. acknowledges support from the MICINN (Spain) through IJC2018-037698-I grant.

\medskip

%

\textbf{References}\\

\end{document}